\documentclass[aps,prl,twocolumn,superscriptaddress,nofootinbib,showpacs]{revtex4-1}
\pdfoutput=1
\usepackage{amsmath,latexsym,amssymb,graphicx,hyperref,color,enumerate}
\usepackage{slashed}

\hypersetup{colorlinks,citecolor= nicegreen,linkcolor= nicered}
\usepackage{color}
\definecolor{nicered}{rgb}{0.7,0.1,0.1}
\definecolor{nicegreen}{rgb}{0.1,0.5,0.1}

\newcommand{\beq}{\begin{equation}}
\newcommand{\eeq}{\end{equation}}
\newcommand{\bea}{\begin{eqnarray}}
\newcommand{\eea}{\end{eqnarray}}
\newcommand{\Tr}{\text{Tr}}

\newcommand\SEC[1]{\smallskip\medskip\noindent{\sf\bfseries #1}}
\newcommand\SUBSEC[1]{\medskip\noindent{\itshape #1}}

\begin{document}
\addtolength{\belowdisplayskip}{-.2ex}       
\addtolength{\abovedisplayskip}{-.2ex}       

\title{Type II Seesaw at LHC: the Roadmap}

\author{Alejandra Melfo}
\affiliation{Universidad de Los Andes, M\'erida, Venezuela}
\affiliation{International Center for Theoretical Physics, Trieste, Italy}
%
\author{Miha Nemev\v sek}
\affiliation{International Center for Theoretical Physics, Trieste, Italy}
\affiliation{J.\ Stefan Institute, Ljubljana, Slovenia}
\author{Fabrizio Nesti}
\affiliation{International Center for Theoretical Physics, Trieste, Italy}
%
\author{Goran Senjanovi\'c}
\affiliation{International Center for Theoretical Physics, Trieste, Italy}
%
\author{Yue Zhang}
\affiliation{International Center for Theoretical Physics, Trieste, Italy}

\date{\today}

\begin{abstract}
\noindent
In this Letter we revisit the type-II seesaw mechanism based on the
addition of a weak triplet scalar to the standard model. We perform a
comprehensive study of its phenomenology at the LHC energies, complete
with the electroweak precision constraints. We pay special attention
to the doubly-charged component, object of collider searches for a
long time, and show how the experimental bound on its mass depends
crucially on the particle spectrum of the theory. Our study can be
used as a roadmap for future complete LHC studies.
\end{abstract}

\maketitle

\SEC{Introduction.}  The modern day understanding of the origin and
the smallness of neutrino mass is based on the see-saw mechanism
\cite{seesaw}. The most natural source for this mechanism is provided
by the Left-Right symmetric theories \cite{leftright}, which require
the existence of the $SU(2)_L$ (and $SU(2)_R$) triplets with
hypercharge $Y=2$. Left-Right symmetry can be realized either at low
scale, or embedded in a grand unified theory such as $SO(10)$. It
turns out that once the see-saw mechanism is turned on, the $SU(2)_L$
triplet gets a small vacuum expectation value, even if it is very
heavy.
%
%
One can even contemplate the possibility that this triplet is the only
low-energy remnant of the new physics beyond the standard model
(SM),\footnote{For instance, in the case of left-right symmetry, it is
  known that the scale must be
  $M_{W_R}\gtrsim2.5\,$TeV~\cite{Maiezza:2010ic} on theoretical
  grounds and 1.7\,TeV~\cite{Nemevsek:2011hz, CMS2} on experimental
  grounds.} in which case one talks of the Type II see-saw
mechanism~\cite{typeII}.

An appealing feature of what could otherwise be seen as an ad-hoc
hypothesis is the minimality and the predictivity of this scenario,
namely, the fact that the Yukawa couplings determine the neutrino mass
matrix. This would become particularly important if the triplet were
to lie in the TeV region, for then its decays could directly probe the
neutrino masses and mixings. 

The doubly charged component of the triplet has been the focus of
attention due to its possibly spectacular signatures at colliders~\cite{Han:2007bk}: if
Yukawa couplings are sufficiently large, it will decay predominantly
into same-sign charged leptons which is a clear signature of Lepton
Number Violation (LNV). The same sign leptons at colliders are a
generic high energy analogue of the neutrinoless double beta decay as
a probe of LNV, envisioned in \cite{KS}.

Both, CDF and D0  performed a search of the doubly charged
component~\cite{Abazov:2011xx}. However, only the pair production of the
doubly charged components was considered. The latest search at
CMS~\cite{CMSDoublySearch} takes into account the associated
production with the singly charged component but assumes the triplet
spectrum to be degenerate. None of them have taken into account the
full complexity of its production and decay modes. An attempt in this
direction was made in~\cite{Akeroyd:2011zz}. Here we provide a
global view of the phenomenological implications of the Type II seesaw
scenario at hadron colliders, in particular at the LHC.

We perform the first electroweak high precision study and demonstrate
the strong dependence of the above CMS limit on the spectrum of the
scalar triplet. In particular we find that the quoted limit on the
order of 250$-$300\,GeV can go down all the way to 100\,GeV for the
mass split around 20$-$30\,GeV.  In what follows we discuss and
quantify our results.

\SEC{The model.}  Let us start by summarizing the salient features of
the Type II see-saw mechanism.  Besides the usual SM particle content,
the model requires the existence of a $Y=2$ $SU(2)_L$ triplet
$\Delta$.
When its neutral component $\Delta^0$ acquires a vev $v_\Delta$, it
generates a Majorana mass for the neutrinos through the Yukawa term
%
\beq 
\label{eqDeltaCouplings} \frac{M_\nu^{ij}}{v_\Delta} L_i^T \, C
i\sigma_2 \Delta L_j + \text{h.c.}\,,
\eeq
where $L_i$ is a left-handed lepton doublet, $C$ the charge conjugation operator and 
\beq \label{eqNuMass}
M_\nu = U^* \, m_\nu \, U^\dagger \,,
\eeq
is the neutrino mass matrix in the basis where the charged lepton masses are diagonal. Here
$m_\nu$ stands for the neutrino masses and $U$ is the PMNS leptonic mixing matrix.
The complete potential for the scalars, including the Higgs doublet $H$, is
\bea
 V &=&  -m_H^2 \, H^\dagger H  + m_\Delta^2 \Tr \Delta^\dagger \Delta    + (\mu H^T \, i\sigma_2 \Delta^* H + \text{h.c.})+
\nonumber \\
 &&{}+ \lambda_1 (H^\dagger H)^2  + \lambda_2 (\Tr \Delta^\dagger \Delta)^2   + \lambda_3 \Tr ( \Delta^\dagger \Delta)^2  +\nonumber \\
 & &{} + \alpha \, H^\dagger H\,  \Tr \Delta^\dagger \Delta  + \beta \, H^\dagger \Delta \Delta^\dagger H \,,
\eea 
and the triplet vev is $v_\Delta = \mu \, v^2/\sqrt 2 \, m_\Delta^2$,
where $v$ is the SM Higgs vev.  Thus a small $v_\Delta$ is technically
natural, as its size is controlled by the $\mu$ parameter which is
only self-renormalized.  A non-vanishing $v_\Delta$ spoils the $\rho$
parameter, which requires $v_\Delta $ smaller than a few GeV.


The triplet components then follow the sum rules  
\begin{gather} \label{eqSumRule1}
m_{\Delta^{+}}^2 - m_{\Delta^{++}}^2 \simeq m_{\Delta^{0}}^2 - m_{\Delta^{+}}^2 \simeq  \beta \, v^2/4\,,
\\ \label{eqSumRule2}
m_S \simeq m_A = m_{\Delta^0}\,,
\end{gather}
where $m_S$ and $m_A$ are the masses of the scalar (S) and
pseudoscalar (A) components of $\Delta^0$.  The triplet components are
separated by equal mass square difference, and there is an upper limit
on the splitting from the perturbativity of $\beta$.
These rules are valid up to tiny $\mathcal O (v^2_\Delta/v^2)$ corrections. 

We first focus on smaller values $v_\Delta \lesssim 10^{-3}\,$GeV,
relevant for probing the connection with neutrino masses at LHC and
later on comment on larger $v_\Delta$ and quantify its upper bound.

\SUBSEC{Probing the flavor structure.}  The doubly charged scalar
$\Delta^{++}$ plays a central role in the physics of this model. In
particular, its decays into same-sign charged leptons probe the
neutrino masses and mixings. This is clear from
\eqref{eqDeltaCouplings}, and is made explicit in the decay rate
\begin{equation}
\label{eqDeltaPPDecay}
	\Gamma_{\Delta^{++}\to\ell_i \ell_j} = \frac{m_{\Delta^{++}}}{8 \pi (1+\delta_{ij})} \left| 
	\frac{(U^* \, m_\nu \, U^\dagger)_{ij}}{v_\Delta} \right|^2.
\end{equation}
%
This connection between the collider physics and the low energy
processes has been studied extensively~\cite{Chun:2003ej, perez&han}. 
If this were the only mode, one could probe the Yukawa flavor
structure though branching ratios to different flavor modes.  In
addition, the decay of the singly-charged component $\Delta^+\to\ell_i
\nu$ may also serve as a possible channel to determine the Yukawa
structure.

\medskip

\begin{figure}
	\centerline{\includegraphics[width=.9\columnwidth]{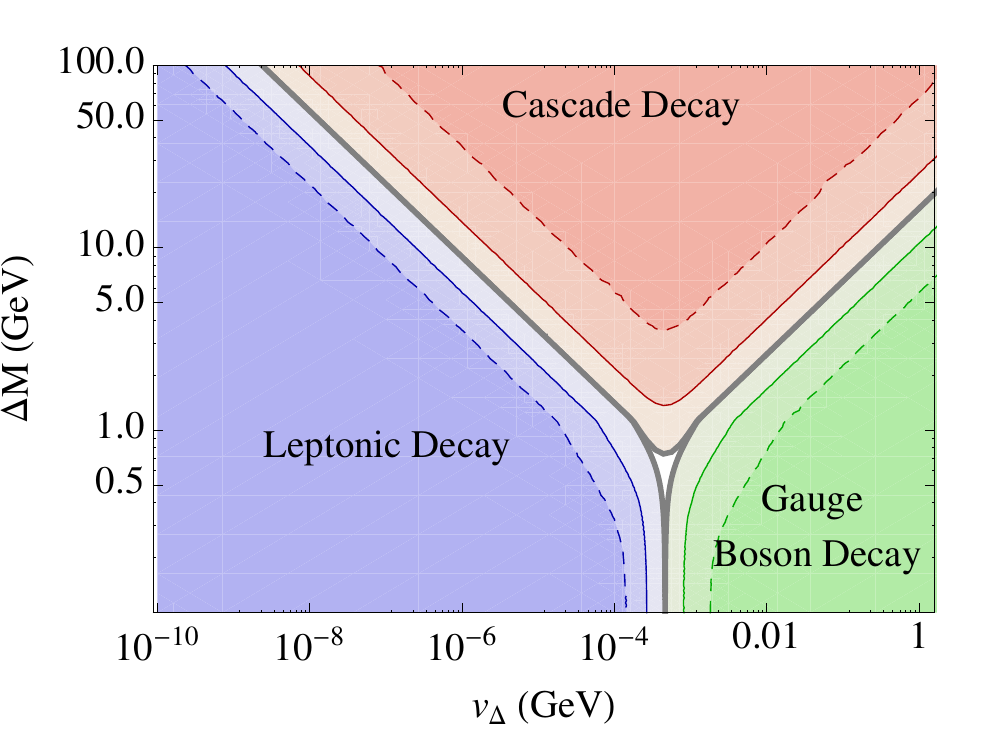}}%
	\caption{Generic decay phase diagram for $\Delta$ decays in the type-II seesaw model, exemplified for
          case B defined in the text, with $m_{\Delta^{++}} =
          150$\,GeV. Dashed, thin solid and thick solid contours correspond to 99, 90 and 50\% of the branching ratios. Here $\Delta M = m_{\Delta^{++}} - m_{\Delta^+}$.
          }
	\label{figBrRegions}
\end{figure}

\SUBSEC{Probing the neutrino mass scale.}  By probing the flavour
structure as above one also measures the ratio of neutrino masses, so
that by using neutrino oscillation data one might infer the absolute
neutrino mass scale. There is also a chance of directly measuring the absolute mass
scale at LHC.  In fact, the other decay
mode, 
%
\begin{equation}
	\Gamma_{\!\Delta^{\!++}\to W^+\!W^+}\! = \frac{g^4v_\Delta^2}{8\pi m_{\Delta^{++}}} \sqrt{1 \!- \!\frac{4 M_W^2}{m_{\Delta^{++}}^2} } \left[ 2\! +\! \left( \frac{m_{\Delta^{++}}^2}{2 M_W^2}\! -\! 1 \right)^{\!2} \right] 
\end{equation}
opens up for a non-vanishing $v_\Delta$. Higgs triplet with gauge
boson fusion production and decay at the LHC has been studied
in~\cite{Godfrey:2010qb}. If large enough this channel would thus
enable the determination of $v_\Delta$.  The critical value is
obtained for $\Gamma_{\Delta^{++} \to \ell_i
  \ell_j}=\Gamma_{\Delta^{++} \to W^+ W^+}$ which gives
$v_\Delta=10^{-4}\div10^{-3}$\,GeV, see Fig.~\ref{figBrRegions}.

\SEC{The decay phase diagram.}  
 The triplet mass sum rules in Eqs.~\eqref{eqSumRule1} and \eqref{eqSumRule2} allow for only two scenarios,
\begin{eqnarray}
&&	{\rm  Case \, A:}\qquad m_{\Delta^0} \geq m_{\Delta^+} \geq m_{\Delta^{++}}  \\[.7ex]
&&	{\rm Case \, B:}\qquad m_{\Delta^{++}} > m_{\Delta^+} > m_{\Delta^0}\,.
\end{eqnarray}
When the triplet components are not degenerate, the cascade channels
$\Delta^0\to\Delta^{+}W^{-*}\to\Delta^{++}W^{-*}W^{-*}$ (for case A)
and $\Delta^{++}\to\Delta^{+}W^{+*}\to\Delta^{0}W^{+*}W^{+*}$ (for
case B) are open~\cite{Akeroyd:2011zz, perez&han}.  These processes
have been overlooked in previous experimental studies due to the
assumption of the degeneracy.

In Fig.~\ref{figBrRegions} we provide a phase diagram separating the
regions where different decay modes play a dominant role. We take as
an example scenario B with $m_{\Delta^{++}} = 150\,$GeV, and consider
the $\Delta^{++}$ decays.  It shows that for moderate mass splits, the
cascade channels become important and one basically looses the
same-sign dilepton channel. Once the mass difference is large enough,
cascade decays quickly dominate.  Similar decaying phase diagrams hold
also for $\Delta^+$ decay in case B and $\Delta^0$, $\Delta^{+}$ in
case A.  On the other hand, for the lightest triplet component there
are only two possibilities: it decays either into leptons or gauge
bosons.
The mass splits have thus a dramatic impact on the direct search
limits on the doubly-charged scalar masses, as we show below.

\begin{figure*} [t!]
  \centering
  \includegraphics[width=\columnwidth]{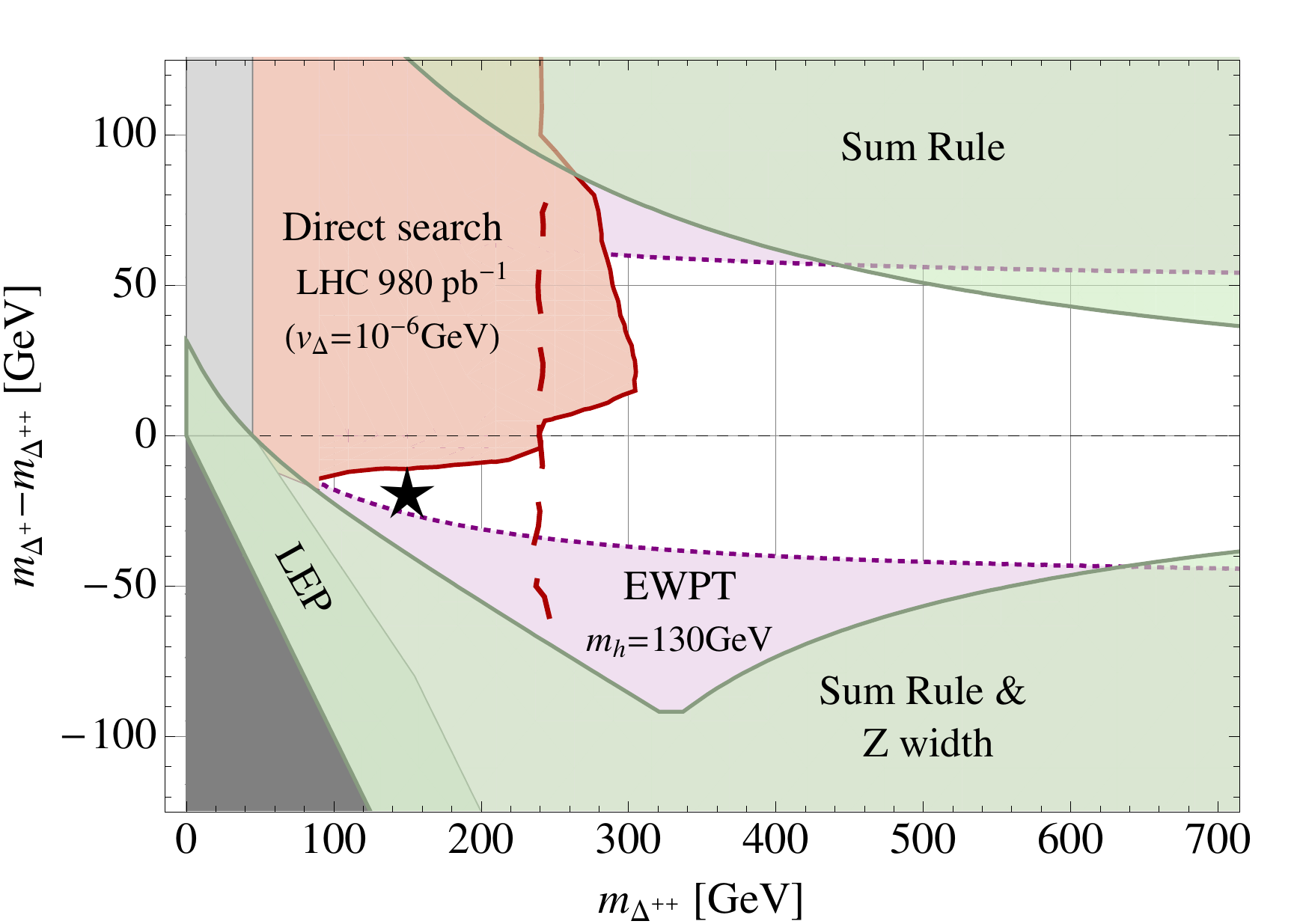}~~ 
  \includegraphics[width=\columnwidth]{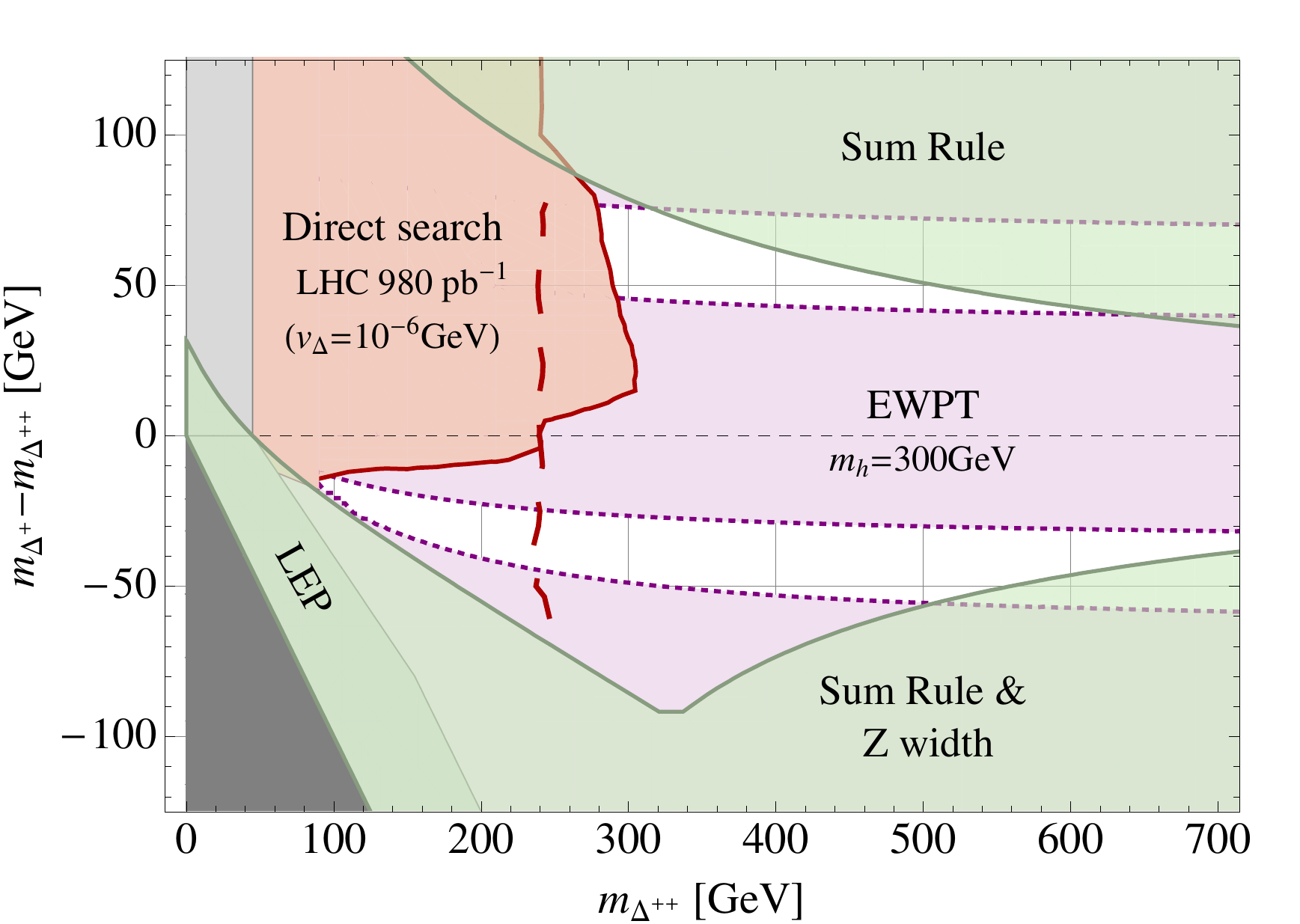}%
\caption{Summary of all the experimental and theoretical constraints
  in the $m_{\Delta^{++}}$--$m_{\Delta^+}$ parameter space, for
  degenerate light neutrino masses. The LHC 2$\sigma$ exclusion is
  shown by the region to the left of the red solid curve, relative to
  $v_\Delta= 10^{-6}$\,GeV.  The analogous curve for $v_\Delta =
  10^{-9}$\,GeV is red dashed. The purple (dotted) contour excluded by
  EWPT at 95\% C.L.  is shown for SM Higgs mass 130\,GeV (left panel)
  and 300\,GeV (right panel). The (green) region excluded by the
  $Z$-width bound and the mass sum rule in Eq.~(4) is shown for the
  triplet-SM Higgs coupling $\beta=3$.  }
  \label{figcollider}
\end{figure*}

\SEC{Electroweak precision tests: a lesson on spectra.} Let us take
this model seriously as an effective theory at the LHC, so that any
other new physics is effectively decoupled. Then, high precision
electroweak study is a must.  We apply the general formulae
in~\cite{Lavoura:1993nq} to the case of the triplet. The dominant
constraint comes from the oblique parameter $T$ which is governed by
the mass differences.
The essential role in this analysis is thus played by the sum rules in
\eqref{eqSumRule1} and \eqref{eqSumRule2}, which eliminate two
arbitrary mass scales. The first message from EWPT is that the mass
split may be large.  In particular, for very light SM Higgs the mass
difference can range from zero to roughly 50\,GeV.  Actually, many of
the studies assumed the degeneracy (or tiny mass difference) among the
members of the triplet.  Although this is possible for a light SM
Higgs, it is strongly disfavored for larger masses, beyond
$~200\,$GeV.  For instance, a very heavy Higgs of 400\,GeV requires the
mass difference to be bigger than $\sim40$\,GeV.  The reason for this
is that the heavy SM Higgs contribution to the $T$ parameter has to be
compensated by a splitting of the triplet components.  There is also
an upper limit on the mass separation due to the sum rule and the
$\beta$ coupling perturbativity, as noted above.  This implies the
triplet mass is bounded from above if SM Higgs boson is heavy.
The above remarks are visible in Fig.~\ref{figcollider} where the
constraints from EWPT and sum rules are brought together with the
collider phenomenology, subject of the next section.

\SUBSEC{$v_\Delta$:~how large?}  Before moving on, let us comment on
the impact of $v_\Delta$ on the EWPT. It simply gives a negative
tree-level contribution to the $T$ parameter: $\Delta T = - 4
v_\Delta^2 / \alpha_{em} v^2$, where $\alpha_{em}$ is the fine structure
constant, and plays a similar role as a heavy Higgs boson (but with
$\Delta S=0$).  The effect of a large $v_\Delta$ can be canceled by a
large mass split, and we find its upper limit from perturbativity
$(\beta \lesssim 3)$ to be $v_\Delta \lesssim7 \,$GeV, for $m_h =
120\,$GeV.

\SUBSEC{$v_\Delta$:~how small?}  A complete study on LFV constraints
has been carried out in~\cite{Akeroyd:2009nu}.  The bottom line is the
combined limit on the vev times the mass of the doubly-charged
component of the triplet
\begin{equation}
v_\Delta m_{\Delta^{++}} \gtrsim 100\,\,{\rm eV} \,\text{GeV}\,.
\end{equation}
These constraints further ensure that the triplet Yukawa couplings are
small enough so that the above EWPT analysis based on oblique
parameters is self-consistent.

\SEC{Current LHC limits.}  The CMS collaboration has published the
latest data on four lepton final states, with a luminosity of
980\,pb$^{-1}$ at $\sqrt s = 7 \text{ TeV}$,
in~\cite{CMSDoublySearch}.  No excess over the SM prediction is
observed and an updated lower limit on the mass of the doubly-charged
Higgs is set.  The analysis is performed assuming degeneracy of the
triplet components. In the following, we perform an estimate of the
limit in the full parameter space.  We generate the events for the
pair and associated production of all the $\Delta$'s using {\tt
  MadGraph\,4.4.57}~\cite{Alwall:2007st}, decay them with {\tt
  BRIDGE\,2.23}~\cite{Meade:2007js} and then do the showering and
detector simulation with {\tt Pythia-PGS
  2.1.8}~\cite{Sjostrand:2007gs, pgs}.  We adopt the K-factor
from~\cite{Muhlleitner:2003me} to account for next-to-leading order
correction to the production. We focus on the four lepton final states
and implement the same cuts as in~\cite{CMSDoublySearch}. These cuts
may be further optimized for different event topologies of cascade
decays, however we would expect only a minor increase of the bound,
due to the rather small triplet splitting. For illustration purposes
we take the triplet vev $v_\Delta=10^{-6}\,$GeV and nearly degenerate
light neutrino masses (corresponding to the sample point BP3
in~\cite{CMSDoublySearch}).

We summarize in Fig.~\ref{figcollider} the limits on the masses of the
charged components, along with the theoretical constraints, i.e.\ the
regions favored by electroweak precision tests at 95\% CL, for SM
Higgs mass of 130\,GeV and 300\,GeV.  The updated lower limit on
$m_{\Delta^{++}}$ for relatively large $v_\Delta$, is independent of
the SM Higgs boson mass.

In case A, we find a lower limit of 240\,GeV on the doubly-charged
Higgs mass for the degenerate case. This is to be contrasted with the
CMS limit of 258\,GeV using four-lepton final states only, probably due
to the use of different statistics.  For moderately large mass splits
this limit can be increased by as much as 50\,GeV, compared to the
degenerate case. We note the analysis can be further improved by
combining both the three- and four-lepton final states, as
done by the CMS collaboration, see also~\cite{Akeroyd:2010ip}.

For case B on the contrary, the limit goes down all the way to
$m_{\Delta^{++}}\gtrsim100\,$GeV (for $v_\Delta > 10$ eV). In this
case, all the $\Delta$ states cascade to $\Delta^0$ and further
to neutrinos. Current missing energy data do not yet possess large
enough luminosity to set here a relevant limit.


We would like to emphasize that: i) the above bounds from CMS data are
valid only for small enough $v_\Delta \lesssim 10^{-4}$\,GeV; ii) the
bounds become splitting independent only for very tiny $v_\Delta$, as
shown by the dashed line with~$v_\Delta=1\,$eV.

\SEC{A look from the right perspective.} As said in the introduction this
scenario can emerge naturally in the context of LR symmetric theories. 
First, the sum rule for $\Delta_L$ remains. Second, $\Delta_R^{\pm}$ gets eaten
by $W_R^{\pm}$, therefore the cascades do not occur and the
limits on $\Delta_R^{++}$ mass set by CDF and D0~\cite{Abazov:2011xx}
remain perfectly valid.

In the LR theory, the neutrino mass situation is more complicated since in general there are both contributions
from type-I and type-II seesaw. In other words, the decay formulae Eq.~(\ref{eqDeltaPPDecay}) gets simply 
modified by the right-handed neutrino masses, mixings and the right-handed triplet vev.
Nonetheless, as long as the competition between the decays into charged leptons and two $W$ bosons exists,
our conclusion on the $m_{\Delta^{++}}$ limit obviously holds true.
Actually, the same conclusion applies in any theory with such phenomena.

\SEC{Implications for the SM Higgs search.}  The crucial couplings to probe in the
Higgs potential are those between the Higgs doublet and the
triplet. For instance the $\beta$ parameter is responsible for the
splitting of the triplet masses, while in a certain region of the
Higgs mass, the $\alpha$ and $\beta$ couplings can be probed through
the Higgs decays to $\Delta$'s~\cite{Akeroyd:2011ir}.

\begin{figure}
\centerline{\includegraphics[width=.9\columnwidth]{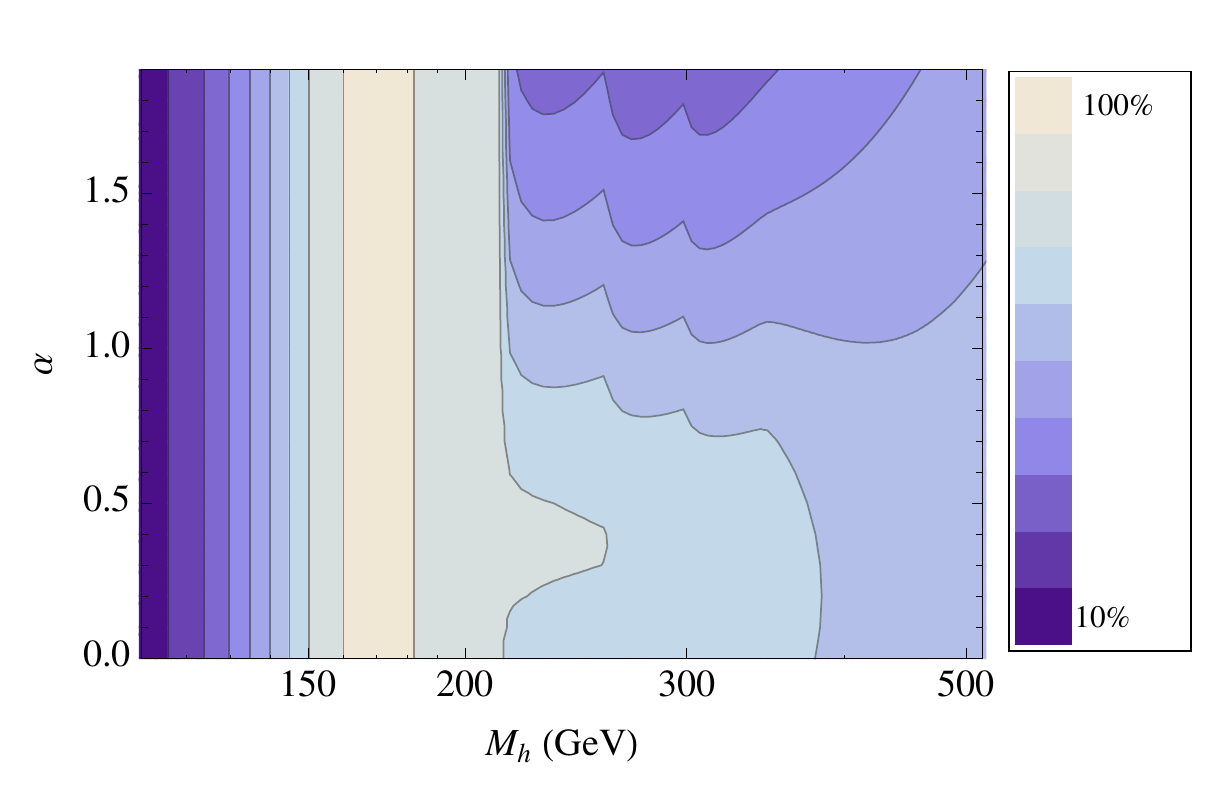}}
\caption{SM Higgs to $WW$ branching ratio for
  $m_{\Delta^{++}}=150$\,GeV and $m_{\Delta^{+}}=130$\,GeV
  (represented by {$\bigstar$} in Fig.~\ref{figcollider}).\label{figHBrvsMh}}
\end{figure}

%
As is well known, a heavy SM Higgs is inconsistent with EWPT, unless
there is new physics near the electroweak scale. In the context of the
type II seesaw, this implies large splits between the components of
the triplet.  When the Higgs is heavier than twice the triplet mass,
the $h\to \Delta\Delta$ channel opens up and may affect the other
branching ratios appreciably. As shown in Fig.~\ref{figHBrvsMh}, the
branching ratio of SM Higgs decay to $W^+W^-$ could be reduced for SM
Higgs heavier than 200\,GeV, and the current limits from the Higgs
search at hadron colliders should be modified.  Interestingly, the
decay to doubly-charged components can in turn serve as another clean
discovery channel for the SM Higgs boson. The opposite case with Higgs
decaying into neutral components with the invisible width controlled
by $\alpha$ could easily explain recent evidence for $m_h \approx 144
\text{ GeV}$.

%
%


\SEC{What next?}  In this letter, we offered a systematic study of the
collider phenomenology for the type-II seesaw mechanism. We showed how
the recently set LHC limit changes dramatically when one moves away
from the assumed benchmark points.
%
%
%
We believe that our results will be a useful roadmap for future
experimental analysis.  We end with a few suggestions for further
exploration.
%
%
\begin{itemize}
\item The missing energy channels relevant for case B require further
  in-depth study, with more statistics.
\item One could try to probe the larger values of $v_\Delta \simeq
  10^{-4} \div 10^{-2}\,$GeV where the di-lepton decay channels
  give rise to displaced vertices, possibly leading to
  simultaneous visibility of both these and $WW$ decay channels.
\end{itemize}
To close, we believe that our work strengthens further the case for
LHC being also a neutrino machine.

\SEC{Acknowledgements.} We are grateful to Georges Azuelos, Dilip K. Ghosh, Ivica Puljak, Beate Heinemann,
Louise Skinnari and Martina Hurwitz for their interest in our work. We thank the BIAS institute for the warm hospitality
and support. YZ thanks the Aspen Center for Physics for hospitality during the final stages of this work. 

\medskip
\SEC{Note added on $\boldsymbol{h\to\gamma\gamma}$.}  \ \
After this work was submitted for publication, both ATLAS and CMS reported~\cite{hgaga} a tentative evidence of the Higgs boson, with a mass about 126 GeV, at 2-3 $\sigma$ CL. In particular, the $h\to \gamma\gamma$ branching ratio is found to be roughly twice as large as the SM prediction. This feature seems to persist in the combined 7 and 8 TeV dataset~\cite{hgaga2012}. Also, a new paper~\cite{Arhrib:2011vc} appeared discussing the $h\to \gamma\gamma$ branching ratio in the type-II seesaw model.  It claims the compatibility with the experimental result for rather large positive values of the quartic coupling $\alpha\sim\mathcal{O}(1-2)$ or larger, depending the masses of the charged components, $\Delta^{\pm, \pm\pm}$. 

\begin{figure}[t]
\centerline{\includegraphics[width=.9\columnwidth]{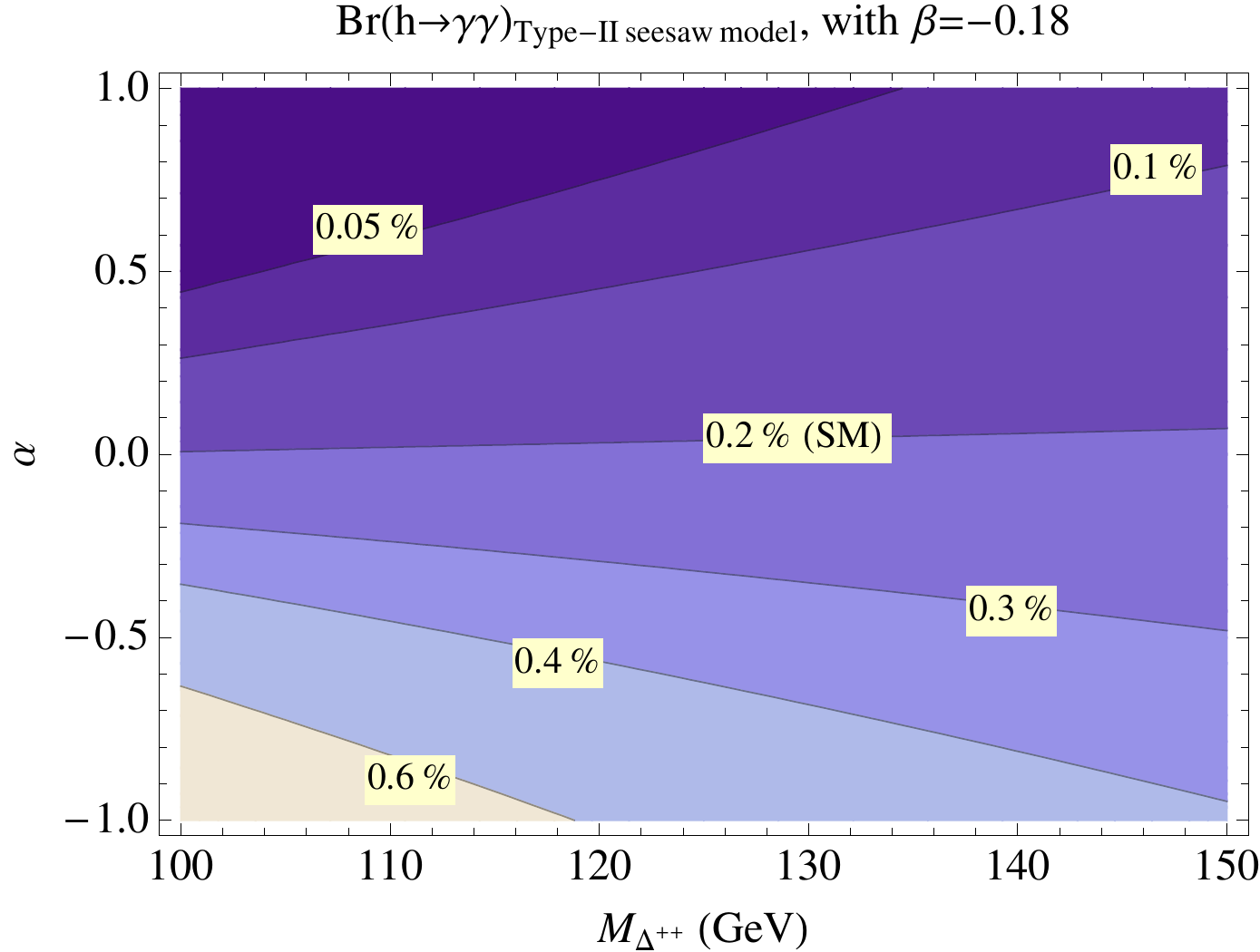}}
\caption{Contours of $Br(h\to\gamma\gamma)$ in the Type II seesaw model, for fixed $\beta=-0.18$. The horizontal contour with $\alpha=0$ is approximately equal to the SM prediction $Br(H\to\gamma\gamma)=\,$0.2\,\%. We find this branching ratio can be enhanced by a factor of 2, for $\alpha\simeq -0.5$ and $m_{\Delta^{++}}\lesssim$\,120~GeV.
\label{hgaga}}
\end{figure}

A new window in agreement with the above LHC results is opened here. As illustrated in Fig.~\ref{hgaga}, a moderate value $\alpha\simeq -0.5$ can do the job, as long as the doubly-charged scalar is light, $m_{\Delta^{++}} \simeq 100 \text{ GeV}$.  This shows how crucial it is to take the cascade decays into account, which is the only way to have such light $\Delta^{++}$, as discussed at length in this paper.

\begin{thebibliography}{99}

%
%
 
\bibitem{seesaw} 
P.~Minkowski,
Phys.\ Lett.\ B {\bf 67} (1977) 421;
R.N.~Mohapatra, G.~Senjanovi\' c,
Phys.Rev.Lett. {\bf 44} (1980) 912;
 S.~Glashow, in {\em Quarks and Leptons, Carg\`ese 1979}, eds.\ 
M. L\' evy. {\em et al.,} (Plenum, 1980, New York);
M.~Gell-Mann, P.~Ramond, R.~Slansky, proceedings of the
{\em Supergravity Stony Brook Workshop}, New York, 1979, 
eds. P. Van Niewenhuizen, D. Freeman (North-Holland, Amsterdam);
T.~Yanagida, proceedings of the {\em Workshop on Unified Theories 
and Baryon Number in the Universe}, Tsukuba, 1979, eds. 
A. Sawada, A. Sugamoto, KEK Report No. 79-18, Tsukuba.

\bibitem{leftright}
J.C.~Pati, A.~Salam,
Phys.\ Rev.\ D {\bf 10} (1974) 275;
R.N.~Mohapatra, J.C.~Pati,
Phys.\ Rev.\ D {\bf 11} (1975) 2558;
G.~Senjanovi\'c, R.N.~Mohapatra,
Phys.\ Rev.\ D {\bf 12} (1975) 1502;
G.~Senjanovi\'c,
Nucl.\ Phys.\ B {\bf 153} (1979) 334.

\bibitem{Maiezza:2010ic}
  G.~Beall, M.~Bander, A.~Soni,
  Phys.\ Rev.\ Lett.\  {\bf 48} (1982) 848;
  for latest studies and further references therein, see A.~Maiezza {\em et al.,} 
  Phys.\ Rev.\  {\bf D82} (2010) 055022;
  Y.~Zhang {\em et al.,} 
  Nucl.\ Phys.\  {\bf B802} (2008) 247.

\bibitem{Nemevsek:2011hz}
  M.~Nemev\v sek {\em et al.,} 
  Phys.\ Rev.\  {\bf D83} (2011) 115014.
  
  \bibitem{CMS2}
The CMS collaboration, {CMS-PAS-EXO-11-002}.

\bibitem{typeII}
 M.~Magg, C.~Wetterich,
 Phys.\ Lett.\  {\bf B94} (1980) 61;
 G.~Lazarides, Q.~Shafi, C.~Wetterich,
 Nucl.\ Phys.\  {\bf B181}  (1981) 287;
 R.N.~Mohapatra, G.~Senjanovi\' c,
 Phys.\ Rev.\  {\bf D23} (1981) 165;
 T.P.~Cheng, L.-F.~Li,
  Phys.\ Rev.\  {\bf D22}  (1980) 2860.


  
\bibitem{Han:2007bk}
  G.~Azuelos, K.~Benslama, J.~Ferland,
  J.\ Phys.\ G {\bf 32} (2006) 73;
 T.~Han {\em et al.,} 
  Phys.\ Rev.\  D {\bf 76} (2007) 075013;
  A.G.~Akeroyd, M.~Aoki, H.~Sugiyama,
 Phys.\ Rev.\  D {\bf 77} (2008) 075010.
  


\bibitem{KS}
  W.-Y.~Keung, G.~Senjanovi\'c,
  Phys.\ Rev.\ Lett.\  {\bf 50 } (1983)  1427; For a review and further references, see
  G.~Senjanovi\' c,
  Riv.\ Nuovo Cim.\  {\bf 034} (2011) 1;
  G.~Senjanovi\' c,
  [arXiv:1012.4104 [hep-ph]].
  
\bibitem{Abazov:2011xx}
  D.~Acosta {\it et al.} [ CDF Collaboration ],
  Phys.\ Rev.\ Lett.\  {\bf 95 } (2005)  071801.
 V.M.~Abazov {\it et al.}  [D0 Collaboration],
 arXiv:1106.4250 [hep-ex],


\bibitem{CMSDoublySearch}
The CMS collaboration, {CMS-PAS-HIG-11-001}, {CMS-PAS-HIG-11-007}.

\bibitem{Akeroyd:2011zz}
  A.G.~Akeroyd, H.~Sugiyama,
Phys.\ Rev.\ {\bf D84} (2011) 035010
and references therein.

%


  
\bibitem{Chun:2003ej}
  E.J.~Chun, K.Y.~Lee, S.C.~Park,
  Phys.\ Lett.\  {\bf B566 } (2003)  142,
 J.~Garayoa, T.~Schwetz,
 JHEP {\bf 0803} (2008) 009;
 M.~Kadastik, M.~Raidal, L.~Rebane,
 Phys.\ Rev.\  {\bf D77} (2008) 115023.
\bibitem{perez&han}
 P.~Fileviez Perez {\em et al.,} 
  Phys.\ Rev.\  {\bf D78 } (2008)  015018.
  

 
\bibitem{Godfrey:2010qb}
  S.~Godfrey, K.~Moats,
  Phys.\ Rev.\  {\bf D81 } (2010)  075026.


  
\bibitem{Lavoura:1993nq}
  L.~Lavoura, L.-F.~Li,
  Phys.\ Rev.\  {\bf D49 } (1994)  1409.


\bibitem{Akeroyd:2009nu}
  A.G.~Akeroyd, M.~Aoki, H.~Sugiyama,
  Phys.\ Rev.\  {\bf D79 } (2009)  113010;
  T.~Fukuyama, H.~Sugiyama, K.~Tsumura,
  JHEP {\bf 1003 } (2010)  044.

\bibitem{Alwall:2007st}
  J.~Alwall {\it et al.,}
  JHEP {\bf 0709} (2007) 028.
   
\bibitem{Meade:2007js}
  P.~Meade, M.~Reece,
  [hep-ph/0703031].
   
\bibitem{Sjostrand:2007gs}
  T.~Sjostrand, S.~Mrenna, P.Z.~Skands,
  Comput.\ Phys.\ Commun.\  {\bf 178 } (2008)  852.
   
\bibitem{pgs}
\url{http://www.physics.ucdavis.edu/~conway/research/software/pgs/pgs4-general.htm}.

\bibitem{Muhlleitner:2003me}
  M.~Muhlleitner, M.~Spira, 
  Phys.\ Rev.\  D {\bf 68} (2003) 117701.

\bibitem{Akeroyd:2010ip}
  A.G.~Akeroyd, C.-W.~Chiang, N.~Gaur,
  JHEP {\bf 1011 } (2010)  005.
  
\bibitem{Akeroyd:2011ir}
  A.G.~Akeroyd, S.~Moretti,
arXiv:1106.3427 [hep-ph].

\bibitem{hgaga}
The ATLAS collaboration, Report No. ATLAS-CONF- 2011-161; 
The CMS collaboration, Report Nos. CMS- PAS-HIG-11-030.

\bibitem{hgaga2012}
The ATLAS collaboration, Report No. ATLAS-CONF-2012-168;
  S.~Chatrchyan {\it et al.}  [CMS Collaboration],
  Phys.\ Lett.\ B {\bf 716}, 30 (2012)
  [arXiv:1207.7235 [hep-ex]].


\bibitem{Arhrib:2011vc} 
  A.~Arhrib, R.~Benbrik, M.~Chabab, G.~Moultaka and L.~Rahili,
  JHEP {\bf 1204}, 136 (2012)
  [arXiv:1112.5453 [hep-ph]].

\end{thebibliography}
\end{document}